# Free Space Few-Photon Nonlinearity in Critically Coupled Polaritonic Metasurfaces

Jie Fang[1,#,*], Abhinav Kala[1,#], Rose Johnson[1], David Sharp[2], Rui Chen[1], Cheng Chang[1], Christopher Munley[2], Johannes E. Fröech[1], Naresh Varnakavi[1], Andrew Tang[1], Arnab Manna[2], Virat Tara[1], Biswajit Datta[3], Zhihao Zhou[1], David S. Ginger[4], Vinod M. Menon[3,5], Lih Y. Lin[1], Arka Majumdar[1,2,*]

[1] Department of Electrical and Computer Engineering, University of Washington, Seattle, WA 98195, USA.

[2] Department of Physics, University of Washington, Seattle, WA 98195, USA.

[3] Department of Physics, City College of New York, New York, NY 10031, USA.

[4] Department of Chemistry, University of Washington, Seattle, WA 98195, USA.

[5] Physics Program, Graduate Center, City University of New York, New York, NY 10016, USA.

[*] Corresponding authors. Email: jiefang@uw.edu; arka@uw.edu

[#] These authors contributed equally: Jie Fang, Abhinav Kala.

**Abstract**

Few-photon optical nonlinearity in planar solid-state systems is challenging yet crucial for quantum and classical optical information processing. Polaritonic nonlinear metasurfaces have emerged as a promising candidate to push the photon number down – but have often been hindered by challenges like the poor photon-trapping efficiency and lack of modal overlap. Here, we address these issues in a self-hybridized perovskite metasurface through critical coupling engineering, and report strong polaritonic nonlinear absorption at an ultra-low incident power density of only 519 W/cm$^2$ (2 orders of magnitude lower than the state of art in free-space planar devices), with an estimated photon number of 6.12 per cavity lifetime. Taking advantage of a quasi-bound-state-in-the-continuum design with asymmetry-controlled quality-($Q$)-factor, we systematically examine the $Q$-dependent device nonlinearity and determine the optimal cavity critical coupling condition. With the optimized device, we demonstrate at 6 Kelvin a tunable nonlinear response from reverse saturable absorption to saturable absorption at varying pump powers, with a maximal effective nonlinear absorption coefficient up to 29.4±5.8 cm/W (6 orders of magnitude larger than unpatterned perovskites) at 560 nm wavelength. In addition, the cavity-exciton detuning dependent device response is analyzed and well explained by a phase-space-filling model, elucidating the underlying physics and the origin of giant nonlinearity. Our study paves the way towards practical flat nonlinear optical devices with large functional areas and massive parallel operation capabilities.

**Introduction**

Achieving nonlinear optical responses at few photons in planar solid-state devices is a crucial but challenging milestone that could advance many technological goals. For instance, free-space optical neural networks based on engineered planar surfaces have demonstrated the advantages of large space-bandwidth products and high computational throughput, but still face a major obstacle due to the absence of effective nonlinear activation functions at low-power optical signals[1–3]. Also, photons are a leading contender for quantum communication and quantum computing, offering clean and decoherence-free 'flying qubits'[4,5]. Flat-optics and metasurface-based quantum-optical technologies have shown promise for multi-channel operation[6] and high-dimensional entanglement[7], yet they remain largely constrained by insufficient photon-photon interactions at few photons[6,8]. Therefore, tremendous efforts are continuously devoted to the quest of strong optical nonlinearity in planar devices[9,10].

Polaritonic nonlinear metasurfaces resulting from non-perturbative strong coupling of photonic cavities and excitonic materials are a promising platform to realize strongly interacting photons in the solid state[11–19]. The half-light, half-matter exciton-polaritons feature a small effective mass, fast polariton-polariton scattering, and strong fermionic phase-space filling effect, which give rise to various nonlinear phenomena like second-harmonic generation[11], correlated quantum fluids[12], and Bose-Einstein condensation[13,18,19].

However, pursuing few-photon polaritonic nonlinearity in free-space sub-wavelength-thick metasurface devices faces several unique and critical bottlenecks, including poor photon-trapping efficiency and photon-to-polariton conversion efficiency, as well as insufficient overlap between the cavity mode and excitonic wave-function. Moreover, different from traditional nonlinear optical systems that prioritize getting ever-higher quality ($Q$)-factor for larger nonlinearity[8–10,17,19], a polaritonic hybrid system requires delicate control and optimization in temporal mode confinement ($Q$ values) to maximize the nonlinear effects.

Here, we investigate these issues and provide the optimized solution through cavity critical coupling engineering in a self-hybridized polaritonic metasurface composed of patterned perovskite crystal (Fig. 1a) that supports both a photonic quasi-bound-state-in-the-continuum (quasi-BIC) mode with asymmetry-controlled $Q$-factor and a highly nonlinear excitonic resonance. We find that a moderately high cavity intrinsic-$Q$ ($Q_{int}$) matching the material background dissipation ($Q_{bg}$) — the cavity critical coupling condition — leads to the best photon-trapping efficiency and the strongest device nonlinear response. We achieve strong polaritonic nonlinear absorption at a record-low (for free-space planar devices, see Table 1)[11,20–24] incident power density of only 519 W/cm$^2$ in our polaritonic metasurface, validating our critical coupling engineering strategy. The photon numbers involved are estimated to be 6.12 or less per cavity lifetime. With such remarkable capability, we further demonstrate at 6 K a tunable nonlinear response, transitioning from reverse saturable absorption (RSA) to saturable absorption (SA) at varying pump powers, with a giant effective RSA nonlinear coefficient up to 29.4±5.8 cm/W and a maximal signal modulation depth of ~10.9 dB at 560 nm wavelength. This offers a versatile planar platform for applications such as nonlinear activation functions in optical neural networks.

To elucidate the underlying physics of the giant optical nonlinearity in our polaritonic metasurface, we investigate the device nonlinearity strengths as a function of the cavity-exciton detuning (quantified by Hopfield coefficients) by leveraging the dispersion of cavity mode at different detection angles (momenta). The results are well explained by a phase-space-filling model. Our study not only suggests a self-hybridized polaritonic metasurface approach towards free space few-photon nonlinearity but also provides detailed design guidance for practical flat nonlinear optical devices at low-power operation.

**Results**

**Quasi-BIC perovskite metasurfaces**

Our goal to realize few-photon polaritonic nonlinearity is two-fold. First, we aim to achieve large nonlinearity with only few photons inside the cavity, necessitating a highly nonlinear material platform and efficient photon-to-polariton conversion. Second, in practice we demand a low incident power threshold for effective nonlinear optical response — a low incident photon number outside the cavity. This further requires exceptional photon-trapping efficiency within a sub-wavelength-thick metasurface device. Below, we outline our design strategy.

Symmetry-protected photonic BICs are ideal dark modes with infinite $Q$-factors[25]. Practically, by slightly breaking the symmetry, one can open a radiative channel to transform the dark BIC into high-$Q$, radiative quasi-BIC modes. This approach has eased the realization of high-$Q$ resonances in free-space meta-optics[17,19,25–27], and thus inspired many exciting advancements in nonlinear optics[17,19] and cavity quantum electrodynamics[27]. Notably, quasi-BIC designs also offer the advantage of easily and accurately tailoring the cavity $Q_{int}$ by varying a single asymmetry parameter, while preserving the near-field profile, mode volume, and other resonance properties mostly unchanged — an aspect that has been relatively underexplored[27,28].

Building upon this, we employ a rod-type $Q$-tunable quasi-BIC design, as depicted in Fig. 1a, to construct self-hybridized perovskite metasurfaces with varying asymmetry parameters and $Q_{int}$. This accurate control of $Q_{int}$ allows us to investigate the cavity critical coupling condition in cavity-material hybrid systems with material background loss, optimizing the photon-trapping efficiency in our metasurfaces (Fig. 1b), as well as to examine the dependence of device nonlinearity strength on the cavity $Q$-factor.

We apply the self-hybridization strategy, patterning excitonic materials into nanostructures to support both photonic cavity modes and excitonic resonances, for better modal overlap between the cavity mode and excitons. Recently, this approach has facilitated several photon-exciton strong coupling studies[19,27,29–33]. Furthermore, thanks to the significant spatial confinement of the quasi-BIC cavity mode, photons can be tightly trapped inside the patterned perovskite crystal, as shown in the simulated near-field profiles (Fig. 1c). This not only enhances the photon-exciton coupling strength, promoting polariton formation, but also spatially confines the formed polaritons in a limited volume for stronger polariton-polariton interactions.

Halide perovskite materials are renowned for their large exciton oscillator strength, making them an attractive polaritonic platform[18,19,32]. In addition, they can exhibit strong optical nonlinearities[34–38], near-unity photoluminescence (PL) quantum yields[39], and permit cost-efficient solution processing. Consequently, we choose halide perovskites in this study as a light-emitting excitonic medium and a highly nonlinear material. Specifically, we select $FAPbBr_3$ perovskite because of its superb thermal and moisture resistance and outstanding crystal stability[36,40], important for a practical device.

The devices are fabricated by spin-coating $FAPbBr_3$ perovskite onto a pre-patterned $SiO_2$ substrate, followed by PMMA encapsulation on top (see Methods for the details). As shown in Fig. 1a, the square holes patterned on the $SiO_2$ substrate are filled by the perovskite crystals, forming the nano-rod building blocks of quasi-BIC metasurfaces. The designed thicknesses of the perovskite nano-rods and PMMA layer are 60 and 50 nm, respectively, and kept fixed in all our devices. The actual thicknesses and perovskite layer morphology slightly deviate from the design (see Methods and Supplementary Fig. S1), which have been accounted for in all simulations and calculations. As highlighted in Fig. 1a, a quasi-BIC unit cell consists of two asymmetric rods, characterized by the following lateral geometrical parameters: the lengths of two rods $L$ and $L$-$\Delta L$; the width of the rods $W$; the distance between two rods $D$; and the period $P_X = P_Y$. The asymmetry parameter is defined as $\alpha = \Delta L/L$ (between 0 and 1), which controls the cavity $Q_{int}$. A multiplicative scaling factor is applied on all the lateral geometrical parameters to spectrally shift the cavity resonance to match the perovskite exciton wavelength.

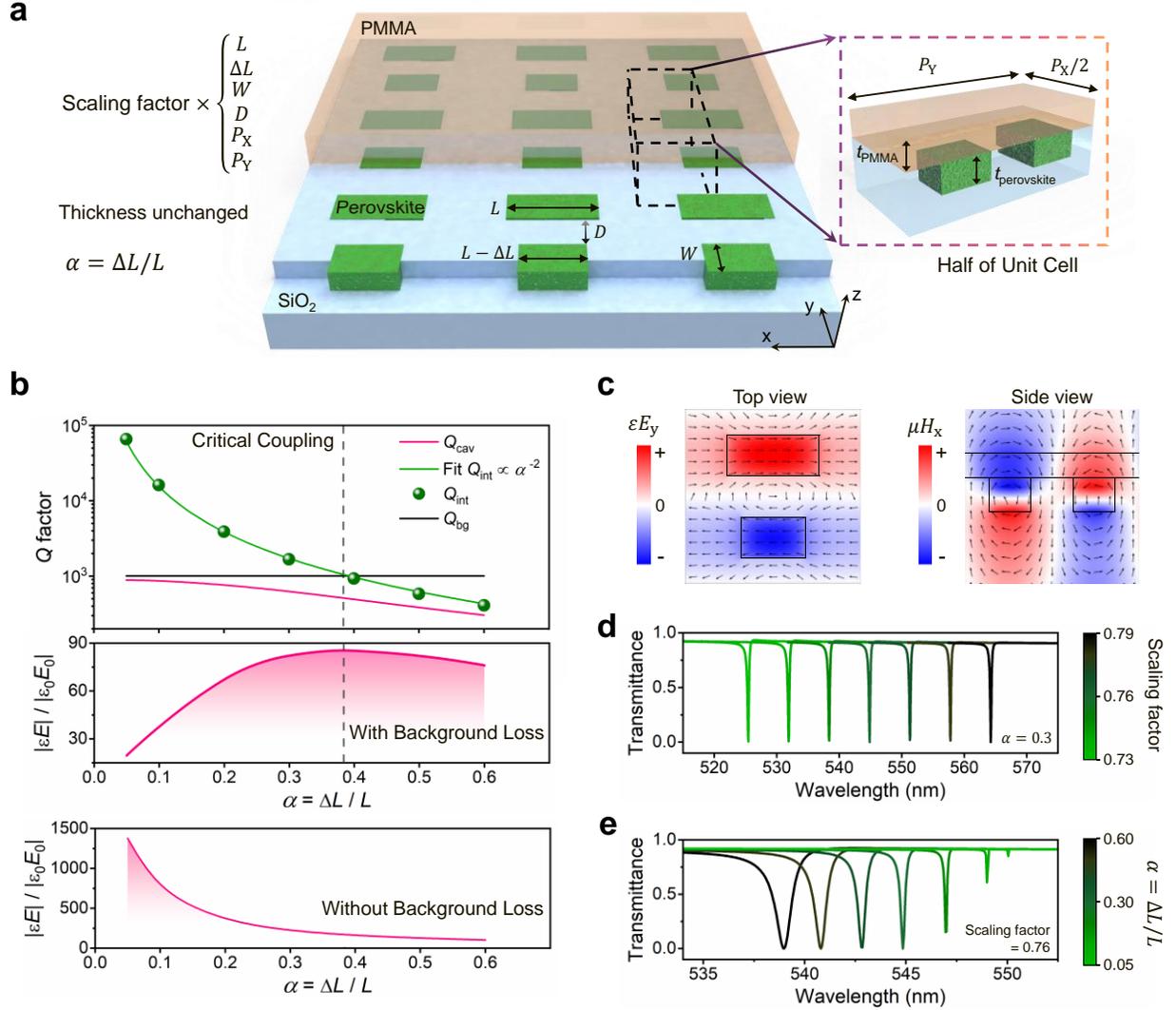

**Fig. 1. Engineering cavity critical coupling in perovskite metasurfaces based on quasi-bound state in the continuum (quasi-BIC).** (**a**) Schematic of the perovskite metasurface in a rod-type symmetry-protected quasi-BIC design. Underneath a PMMA superstrate, a periodic array of $FAPbBr_3$ perovskite nano-rods is embedded in the $SiO_2$ substrate. The geometrical unit cell parameters include the lengths of two asymmetric rods $L = 252.5$ nm and $L - \Delta L = (1 - \alpha)L$, respectively, the width of the rods $W = 112.5$ nm, the distance between two rods $D = 112.5$ nm, the PMMA thickness $t_{PMMA} = 50$ nm, the rod thickness $t_{perovskite} = 60$ nm,

and the period $P_X = P_Y = 475$ nm. An asymmetry parameter $\alpha = \Delta L/L$ between 0 and 1 is applied to tune the cavity intrinsic quality factor $Q_{int}$. A multiplicative scaling factor is applied on the lateral geometrical parameters only (thicknesses unchanged) to tune the resonance wavelength. (**b**) Top, Tuning $Q_{int}$ via different $\alpha$ values to match the material background loss $Q_{bg}$ for the cavity critical coupling. Middle and Bottom, Simulated electric field enhancement inside the perovskite rods as a function of applied asymmetry parameter $\alpha$ when the non-exciton background loss of perovskite is considered (Middle) and not considered (Bottom). A hypothetical perovskite dielectric function with the exciton resonance turned off is applied (see Supplementary Note 1). Scaling factor = 0.76. (**c**) Simulated electric (left) and magnetic (right) field profiles at the quasi-BIC resonance of a perovskite metasurface with $\alpha = 0.30$ and scaling factor = 0.76. The arrows show the directions of field vectors. (**d**, **e**) Simulated transmittance spectra of quasi-BIC perovskite metasurfaces under normal incidence as a function of (**d**) scaling factor and (**e**) asymmetry parameter $\alpha$, respectively. The exciton resonance is turned off.

**Critical coupling for efficient photon-trapping**

An exciton-polariton system consists of a photonic cavity and excitonic Lorentz oscillators from the integrated material. Besides the Lorentz oscillators, solid-state excitonic materials also inevitably involve material background loss. In our analysis, we account for this background dissipation by considering a lossy cavity and aim to optimize its performance.

To begin with, we evaluate the photon-trapping capability as a function of cavity total $Q$-factor, $Q_{cav}$, in an ideal photonic cavity without background dissipation (in other words, $Q_{cav} = Q_{int}$). Considering an incident laser power $P$, a photon-coupling-in efficiency $\eta$ and a detuning $\Delta_{laser}$ between the laser and cavity resonance, the maximum **E** field amplitude inside the cavity can be expressed as[41]

$$|E_{max}| = \sqrt{\frac{\eta P Q_{cav} \lambda_0}{2\pi c \varepsilon V_{cav}} \frac{1}{1+(2\Delta_{laser}/\Delta\omega)^2}}, \qquad (1)$$

where $\lambda_0$ is the cavity resonance wavelength, $c$ is the light speed in vacuum, $\varepsilon$ is the permittivity of the medium, $V_{\text{cav}}$ is the cavity mode volume, and $\Delta\omega$ is the cavity linewidth. Eq. 1 suggests that the field enhancement inside the cavity is proportional to $\sqrt{Q_{\text{cav}}}$.

As mentioned above, the quasi-BIC design allows us to tune $Q_{\text{int}}$ at varying asymmetry parameter $\alpha$, while keeping $\eta$ and $V_{\text{cav}}$ roughly unchanged. Therefore, our quasi-BIC metasurfaces provide an ideal platform to investigate the $E$-$Q_{\text{cav}}$ relationship. In our simulation, we decompose the perovskite dielectric function[37,42] using Tauc-Lorentz model, turn off the exciton oscillator in the numerical simulation[27] to capture the clean cavity property, and further hypothetically exclude any remaining background loss to make $Q_{\text{cav}} = Q_{\text{int}}$ (see technical details in Methods and Supplementary Note 1). The simulated near field enhancement in such a lossless cavity at varying asymmetry parameter $\alpha$ is shown at the bottom of Fig. 1b. The trapped near fields at mode-exciton overlap regions, i.e., within the perovskite nano-rods, are integrated and averaged. The device scaling factor is fixed at 0.76 (cavity resonance matching the exciton wavelength). Consistent with the prediction from Eq. 1, the near field enhancement shows a clear increase at smaller $\alpha$ (higher $Q_{\text{cav}}$).

Furthermore, based on the relationship $Q_{\text{cav}} = Q_{\text{int}} \propto \alpha^{-2}$ in quasi-BICs[17,19,25–27] (see top panel of Fig. 1b), we re-draw Fig. 1b in Supplementary Fig. S2, plotting the near field enhancement as a function of $\sqrt{Q_{\text{cav}}}$. Notably, the plot reveals a perfect linear relationship, quantitively consistent with Eq. 1. This suggests an ever-increasing photon-trapping efficiency with larger $\sqrt{Q_{\text{cav}}}$.

In practice, however, non-zero background loss is inevitable at the exciton wavelength, besides the excitonic oscillator itself. This loss can arise from defects, poly-crystallinity, and the lower-energy tail of above-bandgap absorption band. When such realistic materials are positioned into the cavity modal field, either through traditional cavity-material integration or self-hybridization, the cavity $Q$-factor is affected by the extra dissipation as below[43]:

$$\frac{1}{Q_{\text{cav}}} = \frac{1}{Q_{\text{int}}} + \frac{1}{Q_{\text{bg}}}, \tag{2}$$

where $Q_{\text{bg}}$ is related to the background dissipation rate. A smaller $Q_{\text{bg}}$ means more background loss. This mathematical relationship is intuitively illustrated in the top panel of Fig. 1b. Detailed discussions and perovskite background dielectric function that determines $Q_{\text{bg}}$ can be found in Supplementary Note 1.

Again, we investigate the $E$-$Q_{\text{cav}}$ relationship based on our $Q$-tunable quasi-BIC metasurfaces, this time including the remaining background loss while disabling only the exciton oscillator in simulation. The results are presented at the middle of Fig. 1b. Interestingly, the $E$-$Q_{\text{cav}}$ relationship is no longer monotonic; instead, the near field enhancement reaches its maximum when $Q_{\text{int}} = Q_{\text{bg}}$. As highlighted by the vertical dashed line, we define this as the cavity critical coupling condition, under which we can achieve the best photon-trapping capability. In our devices, it happens when $\alpha$ is between 0.3 and 0.4. Note that this pertains not to the critical coupling of the entire polaritonic system, but only to the lossy cavity in our analysis.

To better understand this phenomenon, we also analytically study a simplified two-interface Fabry-Perot cavity with a lossy medium, as detailed in Supplementary Note 2. The derivation reveals that the maximum photon-trapping efficiency occurs when the cavity radiative rate equals the nonradiative dissipation rate, providing another perspective for interpretating the condition $Q_{\text{int}} = Q_{\text{bg}}$.

Finally, in Supplementary Fig. S3, we tune the material background loss in simulation and compare the corresponding $E$-$Q_{\text{cav}}$ relationships. A cavity critical coupling condition always exists, but the accessible maximum near field enhancement decreases significantly as the loss increases. This decrease highlights the importance of high-purity, low-defect excitonic materials, which set the upper limit for photon-trapping efficiency. Consequently, aiming at few-photon polaritonic nonlinearity, we also optimize the material quality by introducing an antisolvent during the spin-coating of perovskite to accelerate nucleation for better mono-crystallinity (see Methods). The process is extensively tested and optimized, resulting in a high $Q_{bg}$ of ~1035 as shown in Fig. 1b.

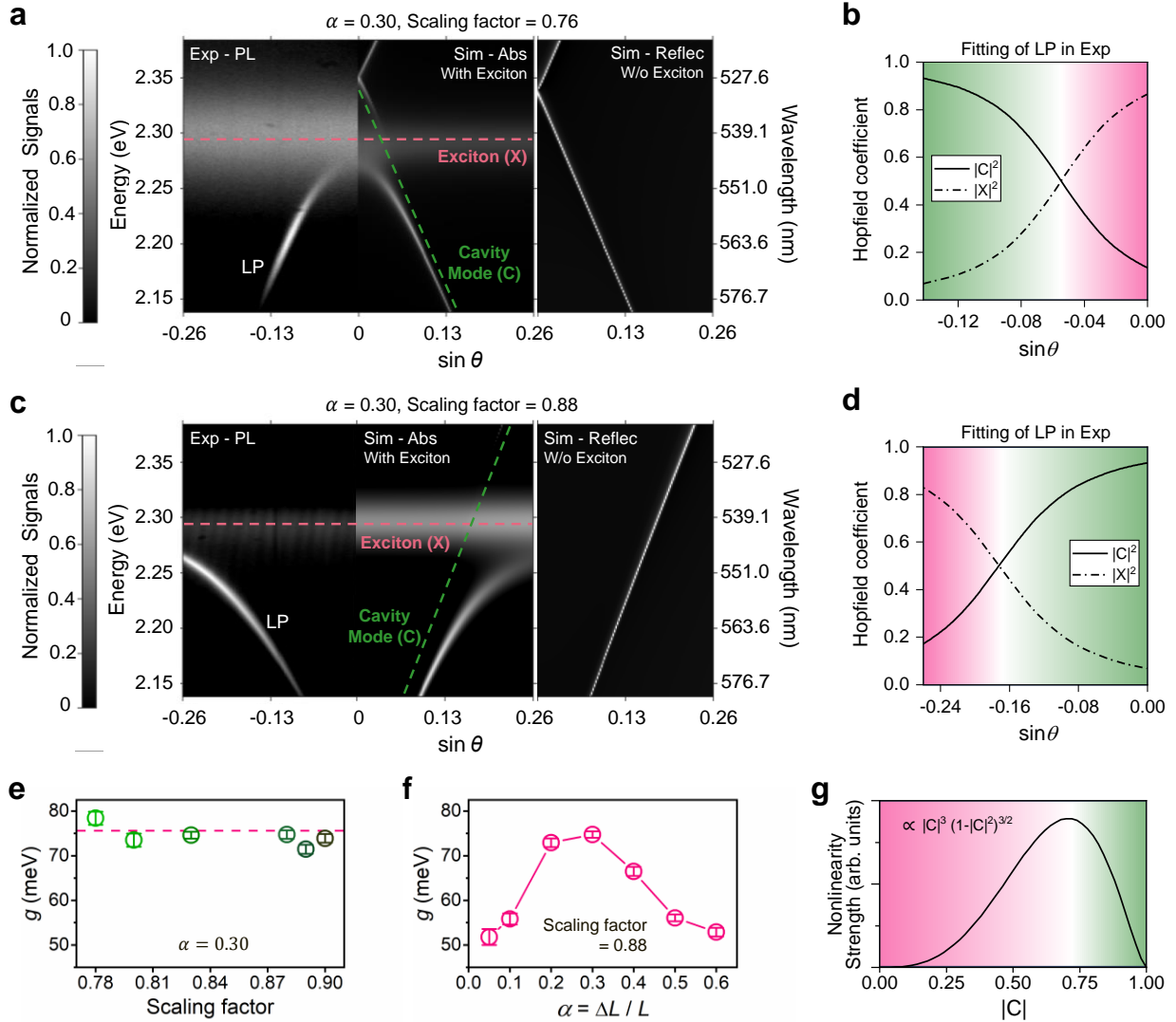

**Fig. 2. Towards large polaritonic nonlinearity under photon-exciton strong coupling.** (**a**) Left, Experimentally measured energy-momentum photoluminescence (PL) spectrum (at 6 K) and simulated absorption (Abs) spectrum of a self-hybridized perovskite metasurface with $\alpha = 0.30$ and scaling factor = 0.76, showing the strong coupling between the photonic cavity mode and perovskite excitons. LP, lower polariton branch. Right, Simulated reflectance (Reflec) spectrum of a hypothetical perovskite metasurface

of the same geometrical parameters but with the exciton resonance turned off (see Supplementary Note 1). (**b**) Fitted Hopfield coefficients as a function of momentum for the studied device in (**a**). A coupled oscillator model is applied to fit the energy-momentum PL spectrum (see Methods and Supplementary Note 3). (**c**, **d**) The same as (a, b) respectively, but for another perovskite metasurface with $\alpha = 0.30$ and scaling factor = 0.88. (**e**) Fitted coupling strengths $g$ of devices with a fixed asymmetry parameter $\alpha = 0.30$ and different scaling factors. (**e**) Fitted coupling strengths $g$ of devices with a fixed scaling factor = 0.88 and different asymmetry parameters $\alpha$, revealing a maximum at the cavity critical coupling condition. (**g**) Theoretical total polaritonic nonlinearity strength as a function of Hopfield coefficient |C| that represents the proportion of photon in the polariton state. |C| can be tuned in the momentum space as shown in (**b**, **d**).

**Self-hybridized exciton-polaritons**

Based on the simulation guidelines in Fig. 1d, 1e, we fabricate a series of devices with scaling factors ranging from 0.70 to 0.89 and asymmetry parameters varying from 0.05 to 0.6 to study the self-hybridized polaritons and their nonlinear optical responses.

Energy-momentum relationship for both reflectance and PL are measured on all samples (see Methods and optical setup schematics in Supplementary Fig. S4). Clear anti-crossing features are observed in Fig. 2a, 2c and Supplementary Fig. S5, S6, confirming the strong coupling and formation of polaritons. Note that the XZ-plane in Fig. 1a serves as the incidence plane and the s-polarization signal is selected at detection. The PL plots in Fig. 2a, 2c are obtained from two metasurfaces with $\alpha = 0.30$ and scaling factors of 0.76 and 0.88, respectively, measured at 6 K. Spectra from other devices and room temperature measurements can be found in Supplementary Fig. S5 and S6, all reaching the strong coupling regime. Unless otherwise stated, all subsequent data presented in the main text are measured at 6 K to narrow the exciton linewidth (increase exciton resonance $Q$-factor), suppressing the cavity-exciton over-coupling when $Q_{cav}$ is larger.

The perovskite dielectric function with the exciton oscillator enabled is used to simulate the momentum-resolved absorption spectra. As shown in Fig. 2a, 2c, and Supplementary Fig. S5, the simulated results exhibit good agreement with the experiments, validating the accuracy of the dielectric function applied for the analysis in Fig. 1. We notice that the upper polariton branch is barely observable in the experimental measurements, consistent with previous reports[19,32]. This is attributed to the strong above-bandgap absorption band at higher energy. In simulations, we artificially suppress this absorption band (see Supplementary Note 1) to reveal the upper polariton spectral profiles. Hypothetical bare cavity dispersive spectral responses are also simulated and depicted on the right side of Fig. 2a and 2c for reference.

To extract the coupling strength, $g$, and momentum-dependent Hopfield coefficients, we employ a coupled oscillator model[44,45] to fit the experimental energy-momentum spectra. The technical details are provided in Methods and Supplementary Note 3. The fitting parameters and fitted $g$ for all the devices are summarized in Supplementary Table S1. For example, the metasurface studied in Fig. 2c ($\alpha = 0.30$, scaling factor = 0.88) exhibits a large coupling strength of 74.7±0.7 meV. The strong-coupling threshold criteria[44,45],

$$g > |\gamma_X - \gamma_C|/2, \tag{3}$$

is well satisfied ($\gamma_X$ =17.2 meV and $\gamma_C$ =2.3 meV are the exciton and cavity linewidths, respectively). Hopfield coefficients $|X|^2$ and $|C|^2$ represent the proportions of excitons and cavity photons in the polariton state, respectively ($|X|^2 + |C|^2 = 1$), and evolve with cavity-exciton detuning. For instance, Fig. 2b (2d) shows the Hopfield coefficient evolution as a function of momentum, extracted from the fittings of the corresponding spectra in Fig. 2a (2c).

Figure 2e and Supplementary Fig. S7 summarize the fitted $g$ as a function of scaling factors with a fixed asymmetry parameter $\alpha = 0.30$. We notice that $g$ increases with the scaling factor up to 0.78 (Supplementary Fig. S7) and then stabilizes with minor fluctuations (Fig. 2e). This trend could be due to the relatively poor perovskite crystallization in smaller holes (patterns with a smaller scaling factor) during spin coating. We therefore avoid this unstable regime by using metasurfaces with a scaling factor of 0.88.

Last, yet most crucially, with a fixed scaling factor of 0.88, we investigate the $Q$-dependent coupling strength $g$ by varying the asymmetry parameter $\alpha$ in Fig. 2f. We observe a trend very similar to that in Fig. 1b, with a maximum at $\alpha = 0.30$. This agreement confirms that maximizing the near field enhancement through critical coupling engineering can also lead to the strongest photon-exciton coupling and thus enhance the polariton formation efficiency. With such a strongly coupled polaritonic platform, we are now one step closer to our goal of few-photon polaritonic nonlinearity. It is also worth noting that all the other devices deviating from the maximum remain well within the strong coupling regime.

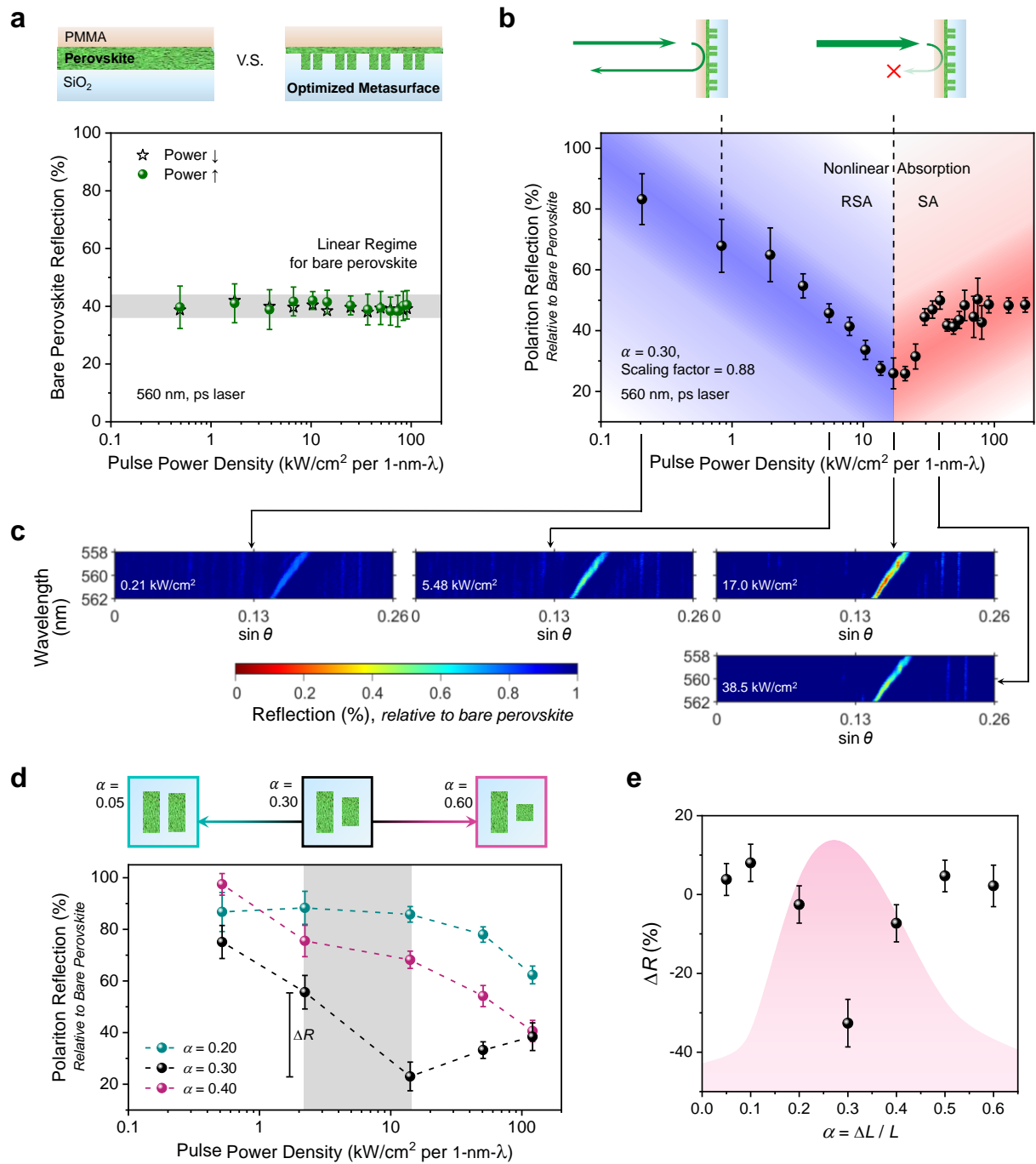

**Fig. 3. Low-photon-number tunable nonlinear absorption of self-hybridized perovskite exciton-polaritons** enabled by cavity critical coupling engineering. (**a**) Power-density-dependent reflection of a bare perovskite thin film on SiO$_2$ substrate and encapsulated by PMMA, showing a repeatable linear optical response. The insets on top are the schematics of a bare perovskite thin film and a patterned perovskite metasurface for comparison in optical response. (**b**) Power-density-dependent relative reflection (compared

to bare perovskite in (**a**)) of polaritons in a self-hybridized perovskite metasurface with $\alpha = 0.30$ and scaling factor = 0.88. The shaded blue and red areas highlight two different regimes with nonlinear optical responses of reverse saturable absorption (RSA) and saturable absorption (SA), respectively. The switching from RSA to SA can be controlled by incident power density. The schematics on top conclude an incident-power-dependent functional switch from 'reflectance allowed' to 'reflectance forbidden', with potential applications in nonlinear optical switches. (**c**) Examples of measured energy-momentum reflection spectra at different power densities, where the polariton reflection data in (**b**) are extracted from. (**d**) Power-density-dependent polariton reflection of different devices deviating from the optimal cavity critical coupling condition, showing the nonlinear optical response as a function of metasurface asymmetry parameters $\alpha$. A parameter $\Delta R$ (grey-shaded area) is defined to evaluate the nonlinearity strength. (**e**) $\Delta R$ as a function of metasurface asymmetry parameters $\alpha$. The pink-shaded background reproduces the trend of fitted coupling strength $g$ as a function of $\alpha$ in Fig. 2f as a reference. The maximum nonlinearity is achieved at the cavity critical coupling condition.

**Giant and tunable polaritonic nonlinear absorption**

With incident photons efficiently trapped and self-hybridized exciton-polaritons established in our optimized perovskite metasurfaces, we proceed to study their nonlinear optical absorption properties by measuring the power-dependent momentum-resolved reflection at polariton wavelengths (Fig. 3c). A 100-picosecond pulsed supercontinuum laser, together with an acousto-optic spectral filter, is employed as a quasi-single-wavelength light source for the on-demand measurements at tunable wavelengths (see Methods). The linewidth of the laser is captured (Supplementary Fig. S8), and only the center 1-nm-wavelength-range power is accounted as effective incident power according to the cavity and polariton linewidths.

Similar to other polaritonic platforms[14,15,46], we notice that the device nonlinearity strength depends on the detuning (Hopfield coefficients). To focus on the dependence of nonlinearity on $Q$-factor only in this section, we exclude other influence by selecting a fixed detuning condition.

Let us assume that the nonlinearity in our system mainly comes from phase-space filling effect[14], and then the detuning-dependent nonlinearity strength should follow the theoretical trend in Fig. 2g. In the next section, we will conduct systematic tests to prove this. Thereby, we choose the 560 nm polariton state in the metasurfaces with scaling factor = 0.88, located around the nonlinearity maximum in Fig. 2g, where $|C|$ = ~0.71.

Under a 560 nm incident laser with pulse power densities ranging from 0.2 to 200 kW/cm², we start the reflection measurements with unpatterned perovskite thin films as a reference (Fig. 3a). The same $SiO_2$ substrate but without patterning is used and the same spin-coating recipe is applied with a PMMA encapsulation on top. Three measurement cycles are performed by first increasing and then decreasing the power, yielding a consistent and repeatable linear optical response in Fig. 3a. The applied power density is insufficient to induce any nonlinear absorption in unpatterned perovskites. The measured reflectance counts from the unpatterned thin film are also utilized to normalize the metasurface reflectance data, enabling a clear presentation of the power-dependent reflection (absorption) change in percentage (Fig. 3b).

Figure 3b shows the power-density-dependent polariton reflection of the optimized perovskite metasurface with $\alpha$ = 0.30 and scaling factor = 0.88. Interestingly, we observe RSA behavior (reverse saturable absorption: a decrease in reflection with increasing power density) in the low power density regime (blue-shaded area) and then it transitions into SA (saturable absorption: an increase in reflection with increasing power density) in the high power density regime (red-shaded area). The transition point is 17.0 kW/cm² and the maximal signal modulation depth (from the highest to lowest reflection signals) is ~10.9 dB, which can be potentially utilized as a nonlinear optical switch as shown in the top schematics in Fig. 3b.

Such RSA-to-SA tunability has also been reported in other halide perovskite materials[38,47]. The RSA originates from bound charge carrier nonlinearity, while the SA is caused by free carrier nonlinearity. As the power density increases, free charges quickly accumulate and become dominant, limiting RSA to the low power density regime only.

Our polaritonic platform exhibits strong nonlinearity even at an ultra-low incident power density and thus the RSA-to-SA transition can be demonstrated. We conservatively consider the average of the two lowest-power data points in Fig. 3b as the operation power threshold for detectable RSA phenomenon, which is only 519 W/cm$^2$. The real operation power could be even lower. This already gives us a record-low (for free-space planar devices) incident power density, 2 orders of magnitude lower than the state of art (see Table 1)[11,20–24]. The corresponding incident photon number outside the cavity is only 2130 per pulse. More excitingly, the photon number per cavity lifetime is estimated to be only 6.12 or less, which represents the effective photon numbers involved in the nonlinear interaction at the same time due to cavity decay.

The photon number calculation is detailed in Supplementary Note 4, and we emphasize that every approximation made in the analysis can only overestimate the photon number. For example, we even consider an idealistic 100% photon trapping efficiency into the cavity. In addition, as mentioned above, we only consider the center 1-nm-range laser power according to the cavity and polariton linewidths. Here, we also report a photon number of ~30.6 per cavity lifetime considering the total laser power for readers' reference.

In Fig. 3d and 3e, we experimentally evaluate the influence of $Q_{int}$ on the polaritonic nonlinear responses based on our quasi-BIC platform ($Q_{int} \propto \alpha^{-2}$). We compare the power-density-dependent reflection of various metasurface devices with asymmetry parameters $\alpha$ ranging from 0.05 to 0.6 (Fig. 3d). For clarity, we present only three representative data sets in the figure, while the plotting of all data sets can be found in Supplementary Fig. S9. The device with $\alpha = 0.30$ exhibits the lowest RSA-to-SA transition power density threshold and the most pronounced reflection change. To quantitatively compare the $Q$-dependent device nonlinearity, we define a parameter $\Delta R$, the difference in reflection between two data points (power densities 2.22 kW/cm$^2$ and 14.1 kW/cm$^2$), as highlighted by the grey-shaded area in Fig. 3d. The measured $\Delta R$ as a function of $\alpha$ is plotted in Fig. 3e.

Interestingly, but also as expected from our cavity critical coupling strategy, $\Delta R$ reaches a maximum absolute value when $\alpha = 0.30$ and quickly drops to near zero when deviating from this optimal condition. In Fig. 3e, we also plot the fitted coupling strength $g$ as a function of $\alpha$ (the same as Fig. 2f) in a shaded background for comparison, which reveals the same trend. These results, particularly the dramatic drop of $\Delta R$ to near zero, significantly highlight the importance of cavity critical coupling in polaritonic nonlinear metasurface devices. Such strong nonlinearity enhancement at the optimal condition arises from the combined effects of enhanced photon-trapping and polariton-formation efficiencies, and the tight spatial confinement of polaritons.

Lastly, to provide a quantitative guideline for practical applications such as nonlinear activation functions in optical computing, we extract the effective nonlinear coefficients for both RSA and SA processes using the following equation:

$$R_{\text{relative}}(I) = \frac{R_{\text{nonlinear}}}{R_{\text{linear}}} = \frac{A_{\text{RSA}} \cdot \exp(-(\alpha_0 + \beta_{\text{RSA}} \cdot I) \cdot L_{\text{eff}}) + A_{\text{SA}} \cdot \exp(-(\alpha_0 + \beta_{\text{SA}} \cdot I) \cdot L_{\text{eff}})}{A_0 \cdot \exp(-\alpha_0 \cdot L_{\text{eff}})}, \quad (4)$$

where $R_{\text{relative}}$ is the presented data in Fig. 3b, the nonlinear reflection signal divided by the linear reflection signal; $I$ is the power density; $A_{\text{RSA}}$, $A_{\text{SA}}$, and $A_0$ are the relative amplitudes of RSA, SA, and linear absorption processes, respectively; $\alpha_0$ is the linear absorption coefficient; $\beta_{\text{RSA}}$ and $\beta_{\text{SA}}$ are the effective nonlinear coefficients for RSA and SA processes, respectively; $L_{\text{eff}}$ is the effective length that light passes through the device/material, which is set to 50 nm (the perovskite thickness) here. More details can be found in Supplementary Note 5.

For the optimal device in Fig. 3b, we get a $\beta_{\text{RSA}}$ of 29.4±5.8 cm/W and a $\beta_{\text{SA}}$ of -7.90±2.06 cm/W. The former is at least 6 orders of magnitude larger than the reported values in nonlinear absorption based on bare halide perovskite materials[34,35,38,40], showing the giant enhancement in nonlinearity strength because of our design and optimization.

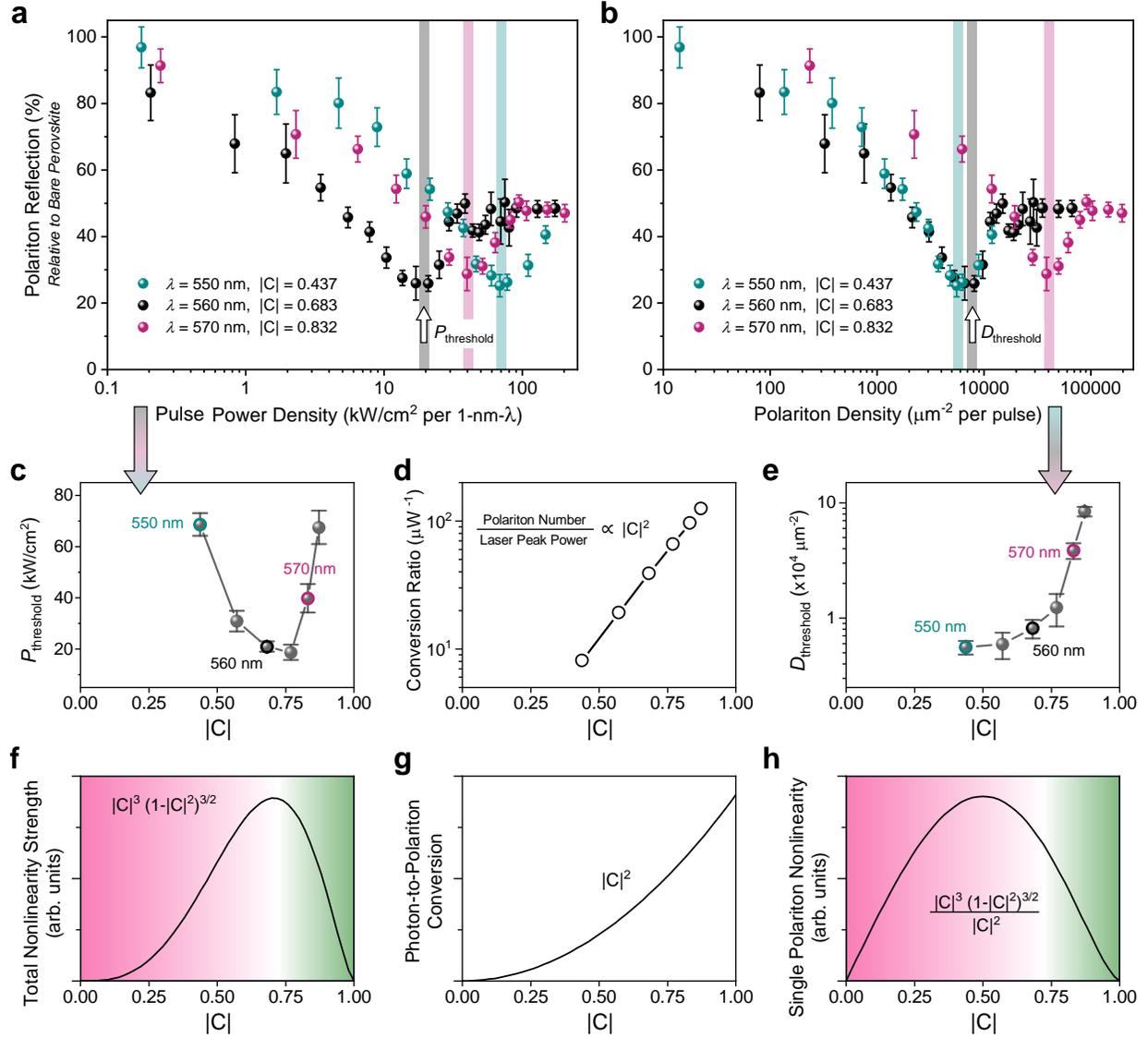

**Fig. 4. Dependence of polaritonic nonlinearity strengths on Hopfield coefficients.** (**a**) Power-density-dependent relative reflection (compared to bare perovskite thin film) of the polaritons of different wavelengths. The same self-hybridized perovskite metasurface with $\alpha = 0.30$ and scaling factor = 0.88 (at cavity critical coupling) in Fig. 3(b) is studied. Along the polariton branch, each polariton wavelength corresponds to a specific momentum value and Hopfield coefficient $|C|$ (see Fig. 2d and Supplementary Fig. S10). (**b**) The same data in (**a**) but the X axis is converted from pulse power density to polariton density. Details of the conversion calculation can be found in Supplementary Note 6. In (**a**, **b**), the transition

thresholds from RSA to SA are defined as $P_{threshold}$ and $D_{threshold}$, respectively. (**c**) $P_{threshold}$ as a function of Hopfield coefficient $|C|$, evaluating the total polaritonic nonlinearity strength. A smaller $P_{threshold}$ suggests more significant nonlinearity. (**d**) Calculated conversion ratio as a function of $|C|$. (**e**) $D_{threshold}$ as a function of $|C|$, evaluating the single polariton nonlinearity strength. A smaller $D_{threshold}$ suggests more significant nonlinearity. (**f** - **h**) Theoretical expectation of (**f**) total polaritonic nonlinearity strength, (**g**) photon-to-polariton conversion ratio, and (**h**) single polariton nonlinearity strength, as a function of $|C|$, according to the phase-space filling mechanism.

**Phase-space filling effect**

To gain deeper insight into the physical origin of the observed polaritonic nonlinearity and to check if the proposed phase-space filling effect is truly dominant, here we study the dependence of nonlinearity strengths on Hopfield coefficients at different cavity-exciton detuning. As shown in Figs. 2c and 2d, the cavity mode dispersion enables momentum-dependent detuning, allowing us to correlate different Hopfield coefficients with the corresponding polariton wavelengths through the momentum relationship (Supplementary Fig. S10). Thereby, with the wavelength-tunable ps laser source, we re-evaluate the optimized device in Fig. 3b at varying wavelengths.

Figure 4a plots the power-density-dependent polariton reflection at different wavelengths (i.e., different Hopfield coefficients). The differences between the data sets are less pronounced compared to those in Fig. 3d, and the RSA-to-SA transition is consistently observed. This is expected, as this time the measurements are performed on the same device, and the detuning does not fundamentally change the cavity critical coupling nature. The most noticeable feature is the variation in the RSA-to-SA transition threshold (the minimum reflection point) among each data set, which is therefore selected as a key metric to quantify the nonlinearity strength. The extracted pulse power density threshold $P_{threshold}$ as a function of Hopfield coefficient $|C|$ is concluded in Fig. 4c. A smaller $P_{threshold}$ suggests a stronger nonlinear interaction.

Then, we compare the trend in Fig. 4c with the $|C|$-dependent total nonlinearity strength $g_{\text{polariton}}^{\text{tot}}$ (Fig. 4f) predicted by the phase-space filling model[14,15]:

$$g_{\text{polariton}} = 4g_{\text{saturate}}|C|(1-|C|^2)^{3/2}, \tag{5}$$

$$g_{\text{polariton}}^{\text{tot}} = g_{\text{polariton}} \cdot N_{\text{polariton}} \propto |C|(1-|C|^2)^{3/2} \cdot |C|^2. \tag{6}$$

In the above equations, $g_{\text{polariton}}$ is the nonlinear interaction strength of single polaritons; $g_{\text{saturate}}$ is the saturated nonlinearity strength, which is a constant; $g_{\text{polariton}}^{\text{tot}}$ is the total nonlinearity strength summing up the collective contribution of all the polaritons; $N_{\text{polariton}}$ is the number of polaritons formed in the system, which is proportional to $|C|^2$ (see the derivation in Supplementary Note 6).

A good agreement is observed between the experiment (Fig. 4c) and theory (Fig. 4f), with the maximum total nonlinear interaction strength occuring at $|C|$ = ~0.71. This can serve as a guide for practical device design, pursuing the lowest operation power. Note that $N_{\text{polariton}}$ is not constant across different incident wavelengths; instead, it also contributes to device performance since a classical device typically relies on the collective effect of all the polaritons.

To further validate the phase-space filling model for our system, we convert the incident power density to polariton density at each incident wavelength (Supplementary Note 6) and evaluate the single polariton nonlinearity as a function of $|C|$. Figure 4b presents the results converted from the data in Fig. 4a. The calculated conversion ratios for each data set are shown in Fig. 4d. As highlighted by the black, cyan, and pink shaded areas in Fig. 4a and 4b, the relative numerical order of the RSA-to-SA thresholds among different data sets (at different detuning) has changed. We extract the new polariton density thresholds $D_{\text{threshold}}$ and plot them in Fig. 4e.

A good agreement is again observed between the experiment (Fig. 4e) and theory (Fig. 4h), regarding the single polariton nonlinearity strength. The maximum now occurs at $|C|$ = ~0.50. This value is also important from a fundamental perspective and could be of interest for applications like quantum information processing, where single-polariton nonlinearity is crucial.

**Discussion**

In conclusion, we have experimentally achieved few-photon polaritonic nonlinearity in self-hybridized perovskite metasurfaces through critical coupling engineering, paving the way towards practical flat nonlinear optical devices with large functional areas and massive parallel operation capabilities.

Based on a quasi-BIC design with asymmetry-controlled $Q$-factor, we investigate the critical coupling condition between the cavity intrinsic $Q_{int}$ and the material background dissipation $Q_{bg}$. Then, through cavity critical coupling engineering, we maximize both photon-trapping and polariton-formation efficiencies in our sub-wavelength-thick polaritonic metasurfaces. The quasi-BIC self-hybridization strategy is applied to facilitate a tightly confined photonic mode and perfect modal overlap; while the optimized $FAPbBr_3$ crystal is chosen to provide a highly nonlinear medium and ensures good mono-crystallization and device stability.

Thanks to all these optimization strategies, we realize strong polaritonic nonlinear absorption at an ultra-low incident power density of only 519 W/cm$^2$ (2 orders of magnitude lower than the state of art in free-space planar devices, see Table 1)[11,20–24], with an estimated photon number of 6.12 or less per cavity lifetime. We demonstrate tunable RSA-to-SA modulation at varying pump powers, achieving a giant effective RSA nonlinear coefficient up to 29.4±5.8 cm/W and a maximal signal modulation depth of ~10.9 dB. In addition, we have studied the evolution of device nonlinearity strength at different detuning and proposed a phase-space filling model that well explain the experimental results, elucidating the origin of giant nonlinearity.

Looking beyond the current cavity critical coupling design with $Q_{int} = Q_{bg}$, further matching $Q_{cav}$ (as defined in Eq. 2) with $Q_{exciton}$ (excitonic oscillator $Q$-factor) could result in a more exquisite polaritonic critical coupling[27,28]. This may lead to even stronger device nonlinearity and further push the photon number down. Indeed, we perform our experimental demonstrations at 6 K in order to get a narrower exciton linewidth and thus a larger $Q_{exciton}$, aiming to approach or even exceed $Q_{cav}$ (see more discussions in Supplementary Note 7). However, $Q_{exciton}$ remains smaller than $Q_{cav}$ here and the whole polaritonic system stays in a slightly over-coupled state.

Given that the applied 100 ps pulse width is 2 orders of magnitude larger than our cavity lifetime, we assume a quasi-CW pumping in our experimental demonstration. More discussions can be found in Supplementary Note 4. Notably, the ultra-low pulse power density of 519 W/cm$^2$ in our demonstration is identical to a 16.3 µW amount of power focused on a 2-µm-in-diameter spot, which can be easily achieved using a CW laser. Practical situations might be more complex as the device responses to pulsed and CW pumping could be different. Unfortunately, a 560 nm CW laser is inaccessible in our lab, but our study can inspire research towards CW-compatible or even incoherent-light-compatible planar nonlinear optical devices.

A large effective nonlinear absorption coefficient offers big nonlinear modulation contrast at very thin material thicknesses (i.e., 50 nm in our devices). Although, alternatively, thicker materials (longer interaction lengths) can always generate sufficient nonlinear modulation contrast, the optical signal loss also dramatically worsens with increasing thicknesses. We notice that significant nonlinear absorption even under incoherent ultra-low-power pumping has been achieved in 10-µm-thick organic macromolecular films under ingenious material design[48]. In contrast, our polaritonic nonlinear metasurface is pushing the few-photon limit in a sub-wavelength ultra-confined manner.

Our study not only introduces a self-hybridized polaritonic metasurface approach towards free space few-photon nonlinearity but also offers detailed design guidance for practical flat nonlinear optical devices at ultra-low-power operation. The few-photon nonlinearity in large-area planar devices could open up new possibilities in quantum-optical technologies. The large signal modulation depth and RSA-to-SA tunability can readily be used in nonlinear optical switches or nonlinear activation functions for free-space optical neural networks.

**Table 1.** Reported operation power density thresholds for free-space nonlinear metasurfaces and ultra-thin planar devices.

| Ref. | Operation power density (W/cm$^2$) | Pump λ (μm) | Design / Structure | Nonlinear effect |
|---|---|---|---|---|
| 11 | 16,400 | 10 | Polaritonic metasurface / Multiple quantum well itself forming a plasmonic nanocavity | Second-harmonic generation |
| 20 | 15,000 | 16 | Hybrid metasurface / Multiple quantum well integrated with a plasmonic metasurface | Second-harmonic generation |
| 21 | 27,000,000 | 0.78 | Mie resonator array / Hydrogenated amorphous silicon nanodisk array | RSA nonlinear absorption |
| 22 | 11,000 | 7.65 | Polaritonic Metasurface / Multiple quantum well itself forming a Mie resonator array | Second-harmonic generation |
| 23 | 181,000,000 | 0.52 | Polaritonic Metasurface / Patterned perovskite itself forming a grating metasurface | SA nonlinear absorption |
| 24 | 30,000,000 | 1.34 | Hybrid metasurface / LiNbO$_3$ integrated with a PB phase metasurface | Second-harmonic generation |
| **This work** | **519** | **0.56** | **Polaritonic metasurface / Patterned perovskite itself forming a quasi-BIC resonance structure** | **RSA and SA nonlinear absorption** |

**Methods**

Patterning of SiO$_2$ substrates

The patterning is made on 8 mm × 8 mm fused silica chips. The chips are thoroughly cleaned by sonication in acetone and isopropanol (IPA), each for 2 minutes, and then treated by oxygen plasma with 150 W power for 5 minutes (AutoGlow, Glow Research). Subsequently, a layer of 200-nm-thick positive-tone resist ZEP 520A is spin-coated and annealed under 180 ºC for 3 minutes, followed by coating conductive polymer (DisCharge H$_2$O). The metasurface pattern is defined by a JEOL JBX-6300FS 100kV electron-beam lithography system and subsequent development in amyl acetate for 2 minutes. Then, SiO$_2$ nano-trenches with a depth of ~60 nm are formed by an inductively coupled plasma reactive ion etching (ICP-RIE) process in florine-based gases. Finally, the ZEP 520A resist is striped by immersing in methylene chloride overnight followed by 5-minute sonication and rinsing in acetone and IPA for 2 minutes.

Perovskite synthesis and metasurface fabrication

For FAPbBr$_3$ perovskite synthesis and metasurface fabrication, the following precursor materials are used without further purification and prepared in a nitrogen-filled glovebox: formamidinium bromide (FABr, >98%, Sigma Aldrich); lead(II) bromide (PbBr$_2$, >98%, TCI Chemicals); dimethyl sulfoxide, anhydrous (DMSO, 99.9%, Sigma Aldrich); N,N-dimethylformamide, anhydrous (DMF, 99.8%, Sigma Aldrich); 1,4,7,10,13,16-hexaoxacyclooctadecane (18-crown-6, >99%, Sigma Aldrich); ethyl acetate, anhydrous (EA, >98%, Sigma Aldrich); poly(methyl methacrylate) (PMMA, 120,000 MW, Sigma Aldrich).

The FAPbBr$_3$ precursor solution is prepared with a 1:1 ratio of FABr and PbBr$_2$ in a 0.4M DMSO/DMF mixture (ratio of 4:1 v/v) and first stirred at 60 ºC for 12 hours. 4 mg ml$^{-1}$ concentration of 18-crown-6 is then added into the precursor solution.

The patterned SiO$_2$ substrates are sequentially cleaned with acetone and isopropanol, followed by oxygen plasma for 10 minutes. The FAPbBr$_3$ precursor solution is filtered with a 0.45-µm PTFE filter and then spin-coated onto the patterned substrate at 7000 rpm for 55 s in a nitrogen-filled glovebox. 150 µL of EA is deposited onto the substrate during the spin-coating process to accelerate nucleation (~35 s after spin start). The timing of adding EA has been extensively tested and optimized to ensure the best mono-crystallization. Then, the sample is immediately annealed on a hot-plate at 100 ºC for 60 s. Afterwards, a thin layer of PMMA is spin-coated on top to encapsulate the perovskite nanostructures. The PMMA solution is prepared with 5 mg ml$^{-1}$ concentration in EA. Finally, the whole device is annealed at 100 ºC for 10 minutes.

Optical setup and measurement

Supplementary Fig. S4 illustrates a schematic of the optical setup, where samples are loaded inside a cryostat (CryoAdvance by Montana Instruments). The setup is based on a confocal configuration, with the excitation path located before the beamsplitter $BS_3$ and the collection path after it.

The excitation beam can be introduced into the system through a single mode fiber (SMF). We use an incoherent white light source (Thorlabs SLS302) for linear reflectance measurements (Supplementary Fig. S5), a 445 nm continuous wave laser (MDL-III-445L by CNI) for linear PL measurements (Fig. 2), and a wavelength-tunable supercontinuum laser source (SuperK Fianium laser and SuperK SELECT acousto-optic tunable filter by NKT Photonics) for power dependent, wavelength-resolved nonlinear reflectance measurements (Fig. 3 and 4). The excitation beam can also be introduced through a flip mirror $FM_1$. We use $FM_1$ to send in a pulsed femtosecond Ti:Sapphire laser (Tsunami by Spectra-Physics) for TPA up-conversion emission generation (Fig 5). There is another flip mirror $FM_2$ for sending in the white light for imaging. A 90:10 beamsplitter ($BS_2$) is placed in the path to capture an optical image of the sample on a CMOS camera. After that, the excitation beam passes through a 50:50 beamsplitter ($BS_3$) and is focused onto the sample inside the cryostat using an infinity corrected objective (Mitutoyo Plan Apo Long WD with 20X magnification and 0.42 NA). The signal reflected from the sample travels back, partially transmits through $BS_3$, and is directed to a spectrometer using another beamsplitter ($BS_1$), for wavelength calibration in our measurements.

The fraction of the collected signal from the sample that is reflected by $BS_3$ is used for energy-momentum spectroscopy. This reflected beam first passes through a linear polarizer (P) and an optional optical filter (OF). The optical filter is used when collecting the PL. For linear PL, we use a long pass filter (Semrock TLP01-501) while for TPA emission, a short pass filter (Thorlabs FES0650) is employed. The signal then enters a 4f system ($L_1$, $L_2$) that images the back focal plane of the objective onto a vertical slit. The focal lengths of lenses $L_1$ and $L_2$ are 18 cm ($f_1$) and 10 cm ($f_2$), respectively. After that, the momentum-resolved back-focal-plane image is further resolved in energy/wavelength using a blazed grating (G). This is accomplished using another 4f system comprised of lenses $L_3$ (focal length $f_3 = 20$ cm) and $L_4$ (focal length $f_4 = 15$ cm). The grating has a groove density of 750 grooves/mm and is placed near the focal spot of $L_3$. Finally, the dispersed light in the first diffraction order of the grating is imaged onto an EMCCD camera (Teledyne ProEM-HS 512).

Numerical simulations

The commercial software Ansys Lumerical is used to simulate the normal-incidence transmission spectra (Fig. 1) as well as the momentum-space-resolved absorption and reflectance spectra of the metasurfaces (Fig. 2) using rigorous coupled-wave analysis method (RCWA). The optical near field profile and enhancement (Fig. 1) under normal-incidence excitation are simulated using a commercial finite element method solver, the wave optics module in COMSOL Multiphysics® 6.0. The real-device geometry slightly deviated from the design is applied in all the simulations to better match the experimental results.

The designed geometrical parameters are concluded in Fig. 1. The real-device in-plane geometrical parameters (except thicknesses) are confirmed by SEM measurements (Thermo Fisher Scientific Apreo 1) before spin coatings. The SEM images can be found in Supplementary Fig. S1. The real thickness of each layer of the metasurface is first measured through a contact profilometer (Bruker DektakXT) and an atomic force microscopy (AFM, Bruker Dimension Icon), and then further examined and adjusted by comparing the simulation and experiment results. Finally, we consider 50-nm-thick perovskite rods embedded in the 60-nm-depth holes in the $SiO_2$ substrate, not fully filled, and a 10-nm-thick perovskite layer on the unpatterned-area surface of $SiO_2$ substrate. A 60-nm-thick PMMA superstrate with a refractive index of 1.493 covers the whole device. A schematic of the real-device geometry used in simulation can be found in Supplementary Fig. S1. The $SiO_2$ substrate has a refractive index of 1.46. The modeling of the refractive index of $FAPbBr_3$ perovskite can be found in Supplementary Note 1.

Coupled oscillator model fitting

The anti-crossing spectra data (Fig. 2 and Supplementary Fig. S5) are fitted according to a coupled-oscillator model in order to extract the coupling strengths and Hopfield coefficients:

$$\begin{pmatrix} E_X + i\gamma_X & g \\ g & E_C + i\gamma_C \end{pmatrix} \begin{pmatrix} X \\ C \end{pmatrix} = E_P \begin{pmatrix} X \\ C \end{pmatrix}. \quad (7)$$

Here, $E_X$ and $E_C$ are the uncoupled exciton and cavity resonance energies, respectively, $\gamma_X$ and $\gamma_C$ are the exciton and cavity resonance linewidths (decay rates), respectively, $g$ is the exciton-cavity coupling strength, and $E_P$ is the polariton state energy corresponding to the eigenvalue of the above matrix. X and C construct the eigenvectors and represent the weighting coefficients of excitons and cavity photons in the polariton state, respectively. Solving the eigenvalue problem results in the expression for the upper and lower polariton branch energies, $E_{LP}$ and $E_{UP}$:

$$E_{LP}, E_{UP} = \frac{1}{2}[E_X + E_C + i(\gamma_X + \gamma_C)] \pm \sqrt{g^2 + \frac{1}{4}[E_X - E_C + i(\gamma_X - \gamma_C)]^2}. \quad (8)$$

For the fitting process, $E_X$ and $\gamma_X$ are directly extracted from experimental data from the bare perovskite thin film and treated as constants. $E_C$ and $\gamma_C$ are extracted from the simulation of a hypothetical perovskite metasurface with the exciton resonance turned off (see Supplementary

Note 1) to mimic an uncoupled cavity. Note that, due to the cavity mode dispersion, both $E_C$ and $\gamma_C$ vary as a function of wavevector/momentum. Since $\gamma_C$ does not change much in the relatively small incidence angle range in our study, it is also treated as a constant, and the simulated value at zero detuning (when $E_X = E_C$) is applied. $E_C$ is allowed to vary by ±20% to account for the difference between simulation and real devices. The coupling strength, $g$, as a free parameter, is then determined by the fitting.

The fittings are conducted using the SciPy Python library's nonlinear least squares curve fitting function. The error bars for the fitted parameters correspond to the standard deviation of fitting errors as reported by the curve fitting function. Furthermore, we can get the Hopfield coefficients $|X|^2$ and $|C|^2$ by solving the eigenvalue problem Eq. 7, using the parameters from the fittings and following the normalization condition $|X|^2 + |C|^2 = 1$.

More details and a list of the fitting parameter values as well as the fitting errors (standard deviation values) can be found in Supplementary Note 3 and Supplementary Table S1.

Fano fitting

Fano fittings are performed using a commercial software MagicPlot. In specific, to fit the simulated spectra in Fig. 1, the Fano fitting equation below is written into the software:

$$T_{\text{Fano}}(\lambda) = a \left[ \frac{\left(b + 2(\lambda-\lambda_0)/\Gamma\right)^2}{1+\left(2(\lambda-\lambda_0)/\Gamma\right)^2} c + (1-c) \right], \qquad (9)$$

where $T_{\text{Fano}}$ is the spectrum to be fitted. $a$, $b$, $c$ are constant real numbers. $\Gamma$ and $\lambda_0$ are the full-width-at-half-maximum (FWHM) linewidth and center wavelength of resonances. The $Q$-factor is then determined by $Q = \lambda_0/\Gamma$.

**Data availability**

All relevant data that support the findings are available within this Article and Supplementary

Information. Source data are available from the corresponding authors upon request.

**Acknowledgements**

J.F. thanks Dr. Kan Yao for the insightful discussion. This paper is based predominantly on work conducted at the University of Washington and City University of New York supported by the U.S. National Science Foundation (NSF) Science and Technology Center (STC) for Integration of Modern Optoelectronic Materials on Demand (IMOD) under Cooperative Agreement No. DMR-2019444. In addition, N.V. and L.Y.L. thank the support from U.S. NSF Grants No. CMMI-2227285 for device fabrication and ECCS-2430679 for perovskite material synthesis. Sample fabrication and characterization work was conducted in part at the Washington Nanofabrication Facility / Molecular Analysis Facility, a National Nanotechnology Coordinated Infrastructure (NNCI) site at University of Washington with partial support from U.S. NSF via awards NNCI-1542101 and NNCI-2025489.

**Author contributions**

J.F. and A.Majumdar conceived the idea. J.F. designed the metasurfaces and experiments. R.C., A.T. and V.T. patterned the $SiO_2$ substrates and did the morphology characterizations. R.J., C.C., N.V., D.S.G. and L.Y.L. synthesized, characterized the perovskite materials and fabricated the devices. A.K. and J.F. performed the experiments with the help from C.M., J.E.F., A.Manna. and Z.Z. J.F. conducted the numerical simulations. D.S. and J.F. did the coupled oscillator model fitting. J.F., A.K., B.D. and V.M.M. did the photon number and polariton density calculation. J.F., A.K.

and A.Majumdar interpreted the results together with all the authors. J.F. wrote the manuscript with input from all the authors and supervised the project with A.Majumdar. J.F. and A.K. contributed equally to this work.

**Competing interests**

The authors declare no competing interests.